\pdfoutput=1
\documentclass[sigconf,screen]{acmart}

\AtBeginDocument{%
  \providecommand\BibTeX{{%
    \normalfont B\kern-0.5em{\scshape i\kern-0.25em b}\kern-0.8em\TeX}}}

\setcopyright{acmlicensed}
\acmPrice{15.00}
\acmDOI{10.1145/3468264.3468617}
\acmYear{2021}
\copyrightyear{2021}
\acmISBN{978-1-4503-8562-6/21/08}
\acmConference[ESEC/FSE '21]{Proceedings of the 29th ACM Joint European Software Engineering Conference and Symposium on the Foundations of Software Engineering}{August 23--28, 2021}{Athens, Greece}
\acmBooktitle{Proceedings of the 29th ACM Joint European Software Engineering Conference and Symposium on the Foundations of Software Engineering (ESEC/FSE '21), August 23--28, 2021, Athens, Greece}

\usepackage{amsmath,amsfonts} 
\usepackage{graphicx}
\usepackage{textcomp}
\usepackage{xcolor}
\def\BibTeX{{\rm B\kern-.05em{\sc i\kern-.025em b}\kern-.08em
    T\kern-.1667em\lower.7ex\hbox{E}\kern-.125emX}}

\usepackage[linesnumbered,ruled,vlined]{algorithm2e}
\usepackage{algpseudocode}
\usepackage{multirow}
\usepackage{xspace}
\usepackage{hhline}
\usepackage{url}
\usepackage{fancyvrb}
\usepackage{xcolor}

\usepackage{todonotes}

\usepackage{microtype}
\newcommand{\substepseparator}{\hspace{1cm}}

\begin{document}
	
\title{Code Integrity Attestation for PLCs using Black Box Neural Network Predictions}

\author{Yuqi Chen}
\affiliation{\institution{Singapore Management University}\country{Singapore}}
\email{yuqichen@smu.edu.sg}

\author{Christopher M. Poskitt}
\orcid{0000-0002-9376-2471}
\affiliation{\institution{Singapore Management University}\country{Singapore}}
\email{cposkitt@smu.edu.sg}

\author{Jun Sun}
\orcid{0000-0002-3545-1392}
\affiliation{\institution{Singapore Management University}\country{Singapore}}
\email{junsun@smu.edu.sg}

\begin{abstract}
	Cyber-physical systems~(CPSs) are widespread in critical domains, and significant damage can be caused if an attacker is able to modify the code of their programmable logic controllers~(PLCs). Unfortunately, traditional techniques for attesting code integrity (i.e.~verifying that it has not been modified) rely on firmware access or roots-of-trust, neither of which proprietary or legacy PLCs are likely to provide. In this paper, we propose a \emph{practical code integrity checking solution} based on privacy-preserving black box models that instead attest the \emph{input/output behaviour} of PLC programs. Using faithful offline copies of the PLC programs, we identify their most important inputs through an information flow analysis, execute them on multiple combinations to collect data, then train neural networks able to predict PLC outputs (i.e.~actuator commands) from their inputs. By exploiting the black box nature of the model, our solution maintains the privacy of the original PLC code and does not assume that attackers are unaware of its presence. The trust instead comes from the fact that it is extremely hard to attack the PLC code and neural networks at the same time and with consistent outcomes. We evaluated our approach on a modern six-stage water treatment plant testbed, finding that it could predict actuator states from PLC inputs with near-$100\%$ accuracy, and thus could detect all 120 effective code mutations that we subjected the PLCs to. Finally, we found that it is not practically possible to simultaneously modify the PLC code and apply discreet adversarial noise to our attesters in a way that leads to consistent (mis-)predictions.
\end{abstract}



\maketitle

\section{Introduction}\label{sec:introduction}

Cyber-physical systems~(CPSs), characterised by their deeply intertwined software components and physical processes, are prevalent in many highly sensitive domains. They appear, for example, as the industrial control systems of critical public infrastructure, automating complex multi-stage processes such as the treatment of raw water or the management of power grids. CPSs of this kind will typically consist of sensors (e.g.~flow indicators) for estimating the state of the physical process, actuators (e.g.~motorised valves) for effecting change on it, as well as programmable logic controllers~(PLCs) that receive sensor readings over a network and compute the appropriate commands to send to actuators. Compromising any one of these components or PLCs can potentially allow an attacker to manipulate the system into a damaging physical state. This has motivated a huge variety of research into defending and assessing CPSs, spanning techniques based on anomaly detection~\cite{Cheng-Tian-Yao17a,Harada-et_al17a,Inoue-et_al17a,Pasqualetti-Dorfler-Bullo11a,Aggarwal-et_al18a,Aoudi-et_al18a,He-et_al19a,Kravchik-Shabtai18a,Lin-et_al18a,Narayanan-Bobba18a,Schneider-Boettinger18a,Carrasco-Wu19a,Kim-Yun-Kim19a,Adepu-et_al20a,Das-Adepu-Zhou20a,Giraldo-et_al18a,Giraldo-et_al20a,Schmidt-Hauer-Pretschner20a}, fingerprinting~\cite{Formby-et_al16a,Gu-et_al18a,Kneib-Huth18a,Ahmed-et_al20a,Yang-et_al20a}, invariant-based monitoring~\cite{Cardenas-et_al11a,Adepu-Mathur16a,Adepu-Mathur16b,Chen-Poskitt-Sun16a,Adepu-Mathur18b,Chen-Poskitt-Sun18a,Choi-et_al18a,Umer-et_al20a,Yoong-et_al21a}, trusted execution environments~\cite{Spensky-et_al20a}, and fuzzing~\cite{Chen-Poskitt-et_al19a,Chen-Xuan-Poskitt-et_al20a,Wijaya-Aniche-Mathur20a}.

Although these techniques successfully cover a number of different CPS threat models, few of the countermeasures specifically address the critical problem of \emph{attesting} the integrity of PLC code, i.e.~verifying that no unauthorised changes have been made to the original program or memory. This can leave the system susceptible to the effects of worms, most famously Stuxnet, which compromised PLCs in nuclear facilities to cause gas centrifuges to tear themselves apart~\cite{Langner11a}. Incorporating an attestation solution can help improve the trustworthiness of devices connected to the physical processes, assuring, for example, that attackers have not discreetly embedded some subtle attack scenario. Unfortunately, traditional attestation solutions cannot readily be applied to the PLCs that CPSs use in practice. Software attestation solutions based on challenge-response protocols~\cite{Castelluccia-et_al09a,Seshadri-et_al04a}, for example, require firmware access to be implemented (which proprietary PLC manufacturers are unlikely to provide), and solutions based on root-of-trust technologies such as TrustZone~\cite{Abera-et_al16a,Alves-Felton04a} or SGX~\cite{Anati-et_al13a} require the use of specific types of hardware from the outset, preventing their use on legacy controllers or low-powered embedded devices.

An alternative approach is \emph{behavioural attestation} (or \emph{CPS attestation})~\cite{Roth-McMillin13a,Valente-et_al14a,Roth-McMillin17a}, in which the integrity of PLC code is attested by comparing its behaviour against mathematical models of how it \emph{should} behave. This is a weaker form of attestation in the sense that that it cannot detect code modifications that do not cause a change of behaviour, but it nonetheless can verify that a PLC is always operating in the expected way. It also has the significant advantages of being possible to deploy without requiring the modification of legacy devices, any access to firmware, or any access to trusted hardware features, which is typically missing from real-world CPSs. Instead, it can be implemented on independent devices (e.g.~Raspberry Pis) that read the inputs/outputs of PLCs, verify them against mathematical models, and alert plant engineers when the actual commands to actuators deviate. These devices should of course be trustworthy too---a problem we address shortly.

The main difficulty in implementing this kind of code integrity checking for PLC programs is in obtaining models that are suitably expressive and accurate for predicting their behaviour, and which can be trained without relying on unrealistic assumptions about the system or attackers. In previous work~\cite{Chen-Poskitt-Sun18a}, we attempted to address this with an approach in which faults (or \emph{mutants}) are injected into the PLC programs and traces of (abnormal) sensor readings are collected. These traces are used to train a supervised machine learning~(ML) model that can effectively classify between behaviour resulting from the original PLC programs and those that have been modified. However, this solution was only assessed on a \emph{simulator} of a CPS, and faces several challenges to be implemented for real PLCs. In particular: (1)~the large amount of data traces it requires would lead to a correspondingly large amount of wasted resources, especially as many random mutations have no physical effects and must be filtered out; (2)~running arbitrarily mutated PLC programs in a real plant is unlikely to be allowed due to safety concerns; and (3)~the solution did not explore the possibility of adversarial attackers with knowledge of the code attestation mechanism.

In this paper, we propose a \emph{practical code integrity checking solution} based on privacy-preserving black box models that attest the behaviour of PLC programs. Our approach can be applied to actual (i.e.~not just simulated) critical infrastructure without incurring any resource wastage or safety issues, and without requiring any trusted hardware components. Instead of mutating code and waiting for physical effects (as in~\cite{Chen-Poskitt-Sun18a}), we mutate the values of PLC inputs and observe the actuator commands that are immediately issued as a result (e.g.~open valve, switch on pump). We implement an information flow analysis to identify the most important of these manipulations, then use the corresponding data to train a highly accurate neural network for predicting actuator commands from PLC inputs, which can thus be used to attest the actual outputs. The practicality and generality of our approach is due the fact that data collection and model training can be done entirely with offline copies of the PLC code that fully characterise the `cyber' part of the system: we avoid the need for any complex (and likely inaccurate) process modelling, eschewing the costs and risks of using the real system during training. Furthermore, our solution exploits the black box nature of the model to ensure that the attester protects the privacy of the original PLC code, and that it remains trustworthy in the presence of attackers who are aware of it and even know the parameters of the model. The trust instead comes from the fact that it is extremely hard to attack PLC code and an ensemble of neural networks at the same time and with consistent outcomes~\cite{Abbasi-Gagne17a,Tramer-et_al18a}.

To evaluate our approach, we implemented our behavioural attestation solution for the Secure Water Treatment~(SWaT) testbed~\cite{SWaT-Reference,Mathur-Tippenhauer16a}, a scaled-down version of a modern six-stage water purification plant. SWaT involves chemical processes such as ultrafiltration, dechlorination, and reverse osmosis, controlled by six PLCs that communicate with each other (and 66 sensors/actuators) over a layered network hierarchy. Using faithful, cross-validated Python translations of its PLC programs, we identified the most important inputs (i.e.~sensor readings, actuator states, and variables), generated data sets relating them to outputs (i.e.~actuator commands), then trained multi-class neural networks for predicting actuator states from different input values. We found them to be effective as attestation models, achieving near-$100\%$ prediction accuracy, and able to detect all 120 of a series of effective PLC code modification attacks. Finally, given that recent work~\cite{Papernot-et_al16a,Wang-et_al19a,Ghamizi-et_al20a,Jia-et_al21a} has shown neural networks to be susceptible to adversarial attacks (i.e.~inputs specially crafted using knowledge of the models in order to cause them to make mistakes), we subjected our solution to an adversarial attacker, finding that it was not practically possible to change the behaviour of PLCs while simultaneously causing the attesters to make consistent (mis-)predictions using discreet adversarial noise.

\substepseparator

\noindent\textbf{Organisation.} In Section~\ref{sec:background}, we present an overview of the SWaT testbed, PLC programs, existing approaches for (remotely) attesting code integrity, and then describe the threat model assumed in our work. In Section~\ref{sec:approach}, we introduce our behavioural attestation solution for PLC code, describing how to implement it for CPSs in general as well as for SWaT specifically. In Section~\ref{sec:evaluation}, we evaluate the effectiveness of our technique for protecting CPSs against code modification attacks and adversarial attacks. In Section~\ref{sec:related_work}, we discuss some related work, before concluding in Section~\ref{sec:conclusion}.

\section{Background}\label{sec:background}

\begin{figure*}[!t]
	\centering
	\includegraphics[width=1\linewidth]{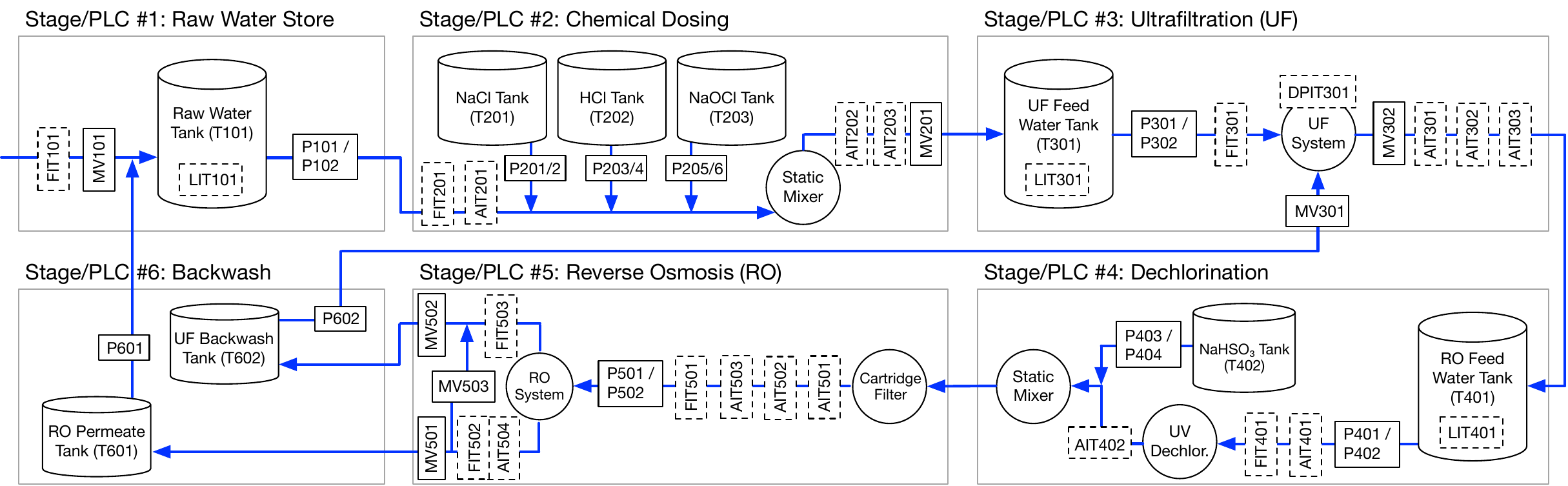}
	\caption{Simplified overview of SWaT (blue arrows indicate water flow; dashed/solid rectangles indicate sensors/actuators)}
	\label{fig:swat_overview}
\end{figure*}

\begin{figure}[!t]
	\centering
	\includegraphics[width=0.8\linewidth]{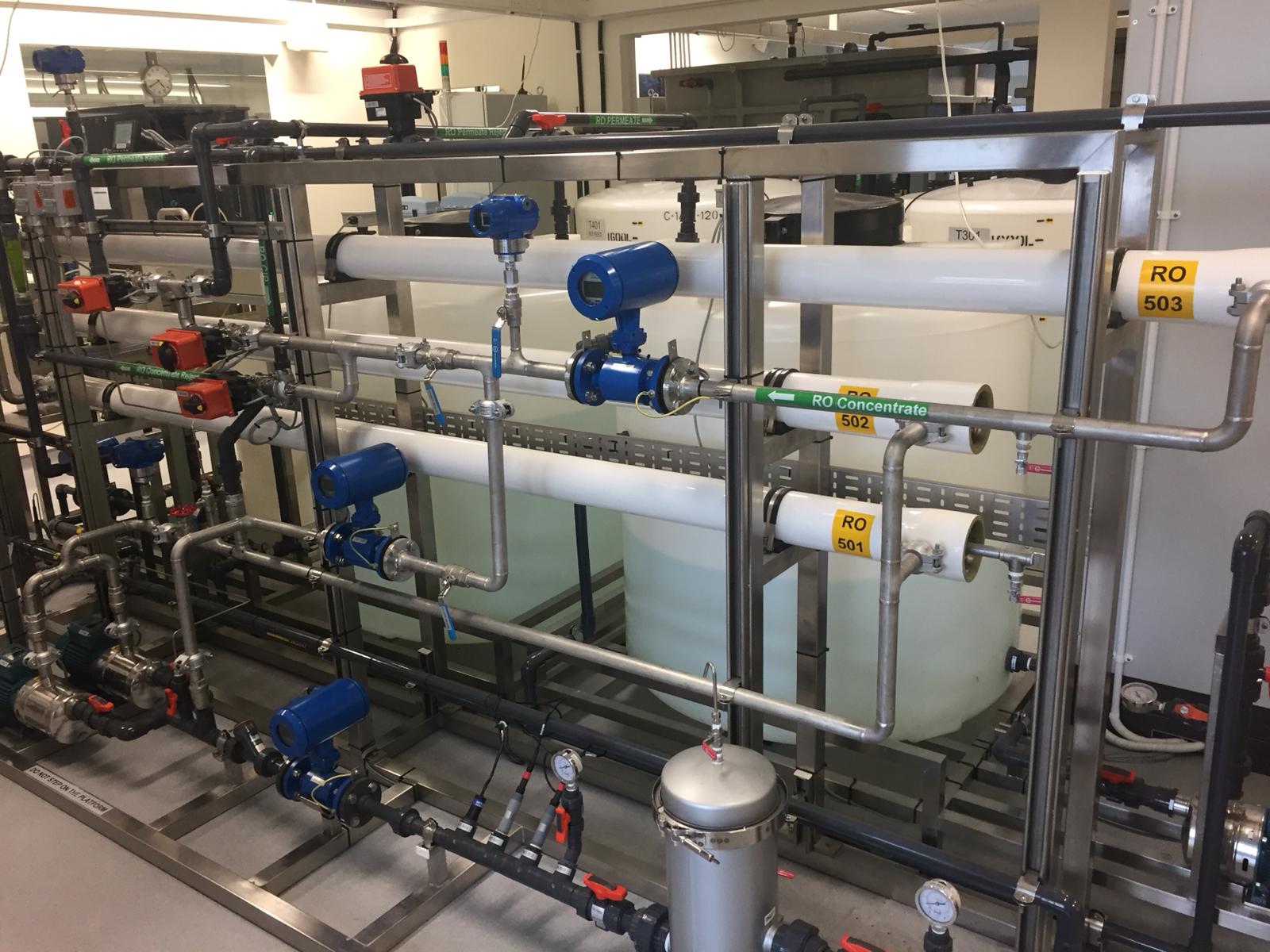}
	\caption{The Secure Water Treatment (SWaT) testbed}
	\label{fig:swat_testbed}
\end{figure}

In this section, we present an overview of SWaT, the water treatment testbed we use to evaluate our approach. Following this, we review existing solutions for remotely attesting code integrity, then give a summary of (and justification for) our threat model.

\substepseparator

\noindent\textbf{SWaT Testbed.} The Secure Water Treatment~(SWaT) testbed~\cite{SWaT-Reference,Mathur-Tippenhauer16a} (Figures~\ref{fig:swat_overview}--\ref{fig:swat_testbed}) is a scaled-down version of a real-world water purification plant, intended to support research into techniques for defending and securing critical infrastructure. SWaT is capable of producing up to five gallons of safe drinking water per minute, which it does so through a six-stage process involving steps such as de-chlorination, reverse osmosis, and ultrafiltration. The testbed has been the subject of multiple hackathons~\cite{Adepu-Mathur18a} involving researchers from both academia and industry. Furthermore, a comprehensive dataset~\cite{Goh-et_al16a} generated from seven days of SWaT's normal operation, as well as four days with attack scenarios, is available for download~\cite{CPS-Datasets} and has been used as a benchmark for evaluating multiple different CPS defence and testing techniques.

The six stages of SWaT are controlled by dedicated Allen-Bradley ControlLogix programmable logic controllers~(PLCs), which communicate with each other as well as the sensors and actuators relevant to each stage. Each PLC cycles through its program, computing the appropriate commands to send to actuators based on the latest sensor readings and values of various \emph{input variables} (of which there are 1747 across the six PLCs). The system consists of 36 sensors in total, including water flow indicator transmitters (FITs), tank level indicator transmitters (LITs), and chemical analyser indicator transmitters (AITs). Among the 30 actuators are motorised valves (MVs) for controlling the inflow of water to tanks, and pumps (Ps) for pumping the water out. Figure~\ref{fig:swat_overview} provides an overview of the six stages, as well as the main sensors and actuators involved.

The network of the SWaT testbed is organised into a layered hierarchy compliant with the ISA99 standard, providing different levels of segmentation and traffic control. The `upper' layers of the hierarchy, Levels 3 and 2, respectively handle operation management (e.g.~the data-logging historian) and supervisory control (e.g.~touch panel, engineering workstation). Level 1 is a star network connecting the PLCs, and implements the Common Industrial Protocol~(CIP) over EtherNet/IP. Finally, the `lowest' layer of the hierarchy is Level 0, which consists of ring networks (EtherNet/IP over UDP) that connect individual PLCs to their relevant sensors and actuators.

The sensors in SWaT are associated with manufacturer-defined ranges of \emph{safe} values, which they are expected to remain within at all times during normal operation. If a sensor reports a (true) reading outside of this range, we say the physical state of the CPS has become \emph{unsafe}. If a level indicator transmitter, for example, reports that the raw water tank in stage one has become more than a certain percentage full (or empty), then the physical state has become unsafe due to the risk of an overflow (or underflow). Unsafe pressure states indicate the risk of a pipe bursting, and unsafe levels of water flow indicate the risk of possible cascading effects in other parts of the system.

\begin{figure*}[!t]
\begin{minipage}{.45\textwidth}
	\centering\scriptsize
\begin{Verbatim}[commandchars=\\\{\}]


2:\textcolor{red}{(*OPEN RAW WATER OUTLET VALVE, MV201*)}
	\textcolor{red}{(*MV-101 , Raw Water Inlet Valve Control*)}
	_MV_101_SR.EnableIn := 1;
	_MV_101_SR.Set := HMI_LIT101.AL;
	_MV_101_SR.Reset := HMI_LIT101.AH;
	\textcolor{blue}{SETD}(_MV_101_SR);  
	_MV101_AutoInp := _MV_101_SR.Out;

	_P_RAW_WATER_DUTY_SR.EnableIn := 1;
	_P_RAW_WATER_DUTY_SR.Set := HMI_MV201.Status =2 AND HMI_LIT301.AL;
	
	_P_RAW_WATER_DUTY_SR.Reset := HMI_MV201.Status <>2 OR HMI_LIT301.AH;
	\textcolor{blue}{SETD}(_P_RAW_WATER_DUTY_SR);
	_P_RAW_WATER_DUTY_AutoInp := _P_RAW_WATER_DUTY_SR.Out;
		
	\textcolor{blue}{IF} HMI_P1_SHUTDOWN \textcolor{blue}{THEN}
			HMI_P1_STATE :=3;
			HMI_P1_SHUTDOWN :=0;
	\textcolor{blue}{END_IF;}
\end{Verbatim}
\end{minipage}\hspace{30pt}
\begin{minipage}{.45\textwidth}
	\centering\scriptsize
\begin{Verbatim}[commandchars=\\\{\}]


2:\textcolor{red}{(*OPEN RAW WATER OUTLET VALVE, MV201*)}
	\textcolor{red}{(*MV-101 , Raw Water Inlet Valve Control*)}
	_MV_101_SR.EnableIn := 0;
	_MV_101_SR.Set := 0;
	_MV_101_SR.Reset := HMI_LIT101.AH;
	\textcolor{blue}{SETD}(_MV_101_SR);  
	_MV101_AutoInp := _MV_101_SR.Out;

	_P_RAW_WATER_DUTY_SR.EnableIn := 1;
	_P_RAW_WATER_DUTY_SR.Set := HMI_MV201.Status =2 AND HMI_LIT301.AL;
	
	_P_RAW_WATER_DUTY_SR.Reset := HMI_MV201.Status <>2 OR HMI_LIT301.AH;
	\textcolor{blue}{SETD}(_P_RAW_WATER_DUTY_SR);
	_P_RAW_WATER_DUTY_AutoInp := _P_RAW_WATER_DUTY_SR.Out;
		
	\textcolor{blue}{IF} HMI_P1_SHUTDOWN \textcolor{blue}{THEN}
			HMI_P1_STATE :=3;
			HMI_P1_SHUTDOWN :=0;
	\textcolor{blue}{END_IF;}
\end{Verbatim}
\end{minipage}
\caption{Structured text program snippet from PLC1 of SWaT before \emph{(left)} and after \emph{(right)} a code modification attack}
	\label{fig:swat_structured_text}
\end{figure*}

SWaT implements a number of standard safety and security measures for water treatment plants, such as alarms (reported to the operator) for when these thresholds are crossed, and logic checks for commands that are exchanged between the PLCs. SCADA software and tools developed by Rockwell Automation are available to allow the plant engineer to monitor and intervene in the system. In addition, several defence mechanisms developed by researchers have been installed~(see Section~\ref{sec:related_work}). Some of these are based on offline analyses, i.e.~that use sensor and actuator data logged by the system's historian.

\substepseparator

\noindent\textbf{PLC Programs.} PLCs are programmed using domain-specific languages that were designed for engineers without a traditional programming background. The IEC 61131-3 open standard~\cite{IECStandard} supports five different high-level languages for PLC programming, including textual languages (e.g.~structured text, instruction list) as well as visual languages (e.g.~ladder diagram, function block diagram). We focus on PLC programs written using the \emph{structured text} language, in which programs are block structured and syntactically similar to Pascal. Roos~\cite{Roos} compares the constructs of Pascal and structured text (e.g.~iteration, case blocks), highlighting the total absence of recursion in the latter. Note that while our paper focuses on one PLC language, the black box nature of our approach allows it to be applied to PLCs programmed in any other.

The example code snippet from PLC1 of SWaT in Figure~\ref{fig:swat_structured_text}~(left) implements the control logic of the raw water inlet valve MV101 and pumps P101/P102 in the process of opening the raw water outlet valve MV201. Specifically, the variable \texttt{MV101\_SR.EnableIn}, when assigned to 1, indicates that the system should open MV101 in this process by default. \texttt{HMI\_LIT101.AL} and \texttt{HMI\_LIT101.AH} are the low and high alarms for the water level of tank T101. (Alarms are critical thresholds for certain sensors in SWaT. If the sensor value reaches a threshold, it will raise the corresponding alarm.) Function \texttt{SETD} is called to check whether some critical conditions are satisfied, which may influence the behaviour of the corresponding actuator. In this example, if the low alarm for LIT101 is triggered, MV101 should be opened, and if the high alarm for LIT101 is triggered, MV101 should be closed. The logic of pump P101/P102 is similar. Pump P101 should be on in this process, while P102 as a standby should be off all the time unless P101 malfunctions. Meanwhile, if MV201 is open and the LIT301 raises a low alarm, pump P101 should be switched on. If MV201 is not open or LIT301 raises a high alarm, then pump P101 should be switched off. Finally, if the system receives the shutdown command, the process of opening MV201 will be over.

The right side of Figure~\ref{fig:swat_structured_text} is the same code snippet but during a code modification attack. In this example, we assume that the attacker maliciously modifies the PLC code to assign the values of \texttt{MV101\_SR.EnableIn} and \texttt{\_MV\_101\_SR.Set} to 0. When the modified code is executed by the system in this process, the valve MV101 will be closed perpetually. As MV101 is the raw water inlet valve, water will not flow into tank T101. As a consequence, tank T101 may underflow while the whole process of water treatment will terminate.

\substepseparator

\noindent\textbf{Remote Attestation.} There are two main categories of techniques for attesting the code integrity of remote (embedded) devices. First, \emph{hardware attestation}, which assumes that the remote device has some trusted platform module~(TPM) or other kind of trust anchor, e.g.~support for ARM TrustZone~\cite{Alves-Felton04a} or Intel SGX~\cite{Anati-et_al13a}. Typically, a verifier issues a challenge to the remote device (the prover), which computes a cryptographic signature or message authentication code over the challenge and a hash of the binary code being attested~\cite{Coker-et_al11a}. It is also possible to remotely attest some dynamic runtime behaviours, e.g.~control flow~\cite{Abera-et_al16a}. Unfortunately, the requirement for specific types of hardware may prevent the use of such techniques in CPSs that use legacy PLCs or low-powered embedded devices (unless they can be interfaced with separate TPM boards~\cite{Salehi-Sarmadi21a}).

Second, \emph{software attestation}, which is intended for resource-constrained devices that are unable to implement schemes based on trusted hardware-level support. These are also challenge-response schemes, but the main difference is that the verifier determines whether or not the prover has been compromised by relying on the amount of time required to compute a response~\cite{Castelluccia-et_al09a,Seshadri-et_al05a,Li-McCune-Perrig11a}. Seshadri et al.~\cite{Seshadri-et_al04a} proposed the first software attestation scheme, which computed a checksum over the memory of the prover in a such a way that an attacker cannot modify that memory without changing the amount of time required to compute the (correct) checksum. To implement such schemes, a procedure must be either pre-programmed into the remote device's memory or downloaded from the verifier. Some recent versions of PLC firmware (e.g.~in the Siemens S7 series) include APIs for checksum generation over control logic which can help support some software attestation schemes~\cite{Ghaeini-et_al19a}, but without this, software attestation may be challenging in general, as PLCs are typically proprietary, and manufacturers may be unwilling to support firmware access.

Given the challenges of applying hardware and software attestation techniques to CPSs, a third category of techniques has emerged in which controllers are attested by comparing their behaviour against mathematical models of how they \emph{should} behave. Valente et al.~\cite{Valente-et_al14a}, for example, propose that verifiers introduce false control signals and compare the system dynamics against a model to attest that sensors and controllers are operating correctly. Roth and McMillin~\cite{Roth-McMillin13a,Roth-McMillin17a}, similarly, use process feedback to implement a distributed attestation protocol for a smart grid. As neither of these \emph{behavioural attestation} schemes require hardware or firmware access, they can be applied to CPSs such as SWaT once a suitably expressive and accurate model has been constructed for predicting the controllers' effects on sensors and actuators (as discussed in Section~\ref{sec:introduction}).

\substepseparator

\noindent\textbf{Threat Model.} We assume that potential attackers are insiders (i.e.~attackers originating within the organisation) who have access to the PLC code, as well as the ability to arbitrarily modify it. As attackers need to be able to check whether their PLC code modifications successfully brought about unsafe physical states, we assume that they are able to access the true sensor readings by some means. Furthermore, we assume that attackers are aware of the presence of PLC attesters, have full knowledge about their underlying models (e.g.~the parameters of the neural networks), and are able to intercept and modify the sensor, actuator, and input variable values received by the PLCs and attesters (note that attesters and PLCs are physically connected on the same network and thus receive the same inputs). These are standard assumptions in the context of attackers that are able to craft adversarial samples~\cite{Papernot-et_al16b}.

In practice, insiders may only have a strict subset of these abilities, but by assessing our PLC attesters against a theoretically powerful attacker, we can conclude that we can also defend against weaker attackers that may exist in reality.

\section{Approach and Implementation}\label{sec:approach}

The overall goal of our approach is to be able to remotely attest that PLCs are \emph{behaving} in the same way as they would under their original programs, without requiring any trusted hardware, without requiring resource-intensive training methods, and without requiring the unrealistic assumption that our verifiers are themselves impenetrable from attack. To achieve this, we propose a practical and general attestation solution based on predictive models for PLC inputs/outputs. Our models can be trained entirely on offline code while remaining highly accurate for the real CPS.

Our approach consists of the following main steps. First, input/output \emph{data collection} using faithful offline copies of the PLC programs, with the most important inputs identified using an information flow analysis. Second, the \emph{training} and validation of neural network models for predicting PLC behaviours. (Algorithm~\ref{alg:overall} provides a high-level overview of these two steps for obtaining a model.) Finally, with the models obtained, their deployment in devices to \emph{behaviourally attest} the PLCs of the real system.

\begin{algorithm}[!t]
\caption{Learning a model for attestation}\label{alg:overall}
\KwIn{PLC $P$; set of labels $L$ encoding all combinations of actuator commands}
\KwOut{Model for attestation $M_P$}
Translate program of $P$ to imperative program~$C_P$ with an ordered vector of inputs $I_{C_P}$;\\
Cross-validate $C_P$ against data traces from $P$;\\
Identify the most important inputs $I_{C_P}'$ of $C_P$ (Alg.~\ref{alg:relevant_variables});\\
$trainData := \{\}$;\\
\Repeat{timeout}
{
	Randomly generate values for all inputs $I_{C_P}$ and run $C_P$ on them;\\
	Let $l\in L$ denote the label encoding the actuator commands issued by $C_P$;\\
	$trainData := trainData \cup \{ \langle [v_0, v_1, \dots v_{n-1}], l \rangle \}$ where the $v_i$s are the values of $I_{C_P}'$;\\
}
Train a model $M_P$ on $trainData$;\\
\Return $M_P$;
\end{algorithm}

In the following, we describe how to implement these three broad steps in general, and how they were implemented for the SWaT water treatment testbed (Section~\ref{sec:background}) in particular.

\subsection{Data Collection by Input Mutation}\label{sec:approach:step1}

In this step, we convert the PLC programs (e.g.~structured text code) to equivalent imperative programs (e.g.~in Python) that can be utilised for data collection offline. Following this, we perform an information flow analysis to identify the most important inputs for a given PLC (i.e.~those most likely to lead to a change in behaviour), and then repeatedly execute the offline version on different values of those inputs to determine what the digital output commands to the actuators would be. The goal of this data collection is to be able to train models on sufficiently large sets of input/output relations so as to cover as many normal and abnormal (e.g.~exception handling) scenarios as possible.

\substepseparator

\noindent\textbf{Offline PLC Code.} The main prerequisite to implement our approach for a CPS is to have faithful offline copies of its PLC programs, written in an imperative programming language. This is a reasonable assumption to make: a PLC program (e.g.~ladder logic, function block diagram, or structured text) largely consists of a sequence of assignments and conditional statements that are executed repeatedly over a continuous loop. This kind of logic is easily translated to the control structures of traditional imperative languages. For CPSs with multiple PLCs, we translate each PLC into a separate script that can be run independently. This assumption can be made as the outputs are determined uniquely by the inputs (PLCs are deterministic). Furthermore, when considering a PLC in isolation, process modelling is not required to compute the output.

\emph{Implementation for SWaT.} We satisfied this prerequisite by adapting the SWaT simulator used in previous work~\cite{Chen-Poskitt-Sun18a} which contains Python translations of the six actual PLC programs (i.e.~faithful translations of the original structured text code such as Figure~\ref{fig:swat_structured_text}). Our adaptation extends the simulator by adding the control logic for chemical processes and pressure---the original version focused only on water flow. We did not require the (partial) process models included in the original simulator, which approximated the physics of water flow. Instead, we use \emph{only} the simulated PLC code, which has been translated accurately and cross-validated against the real PLCs operating in the testbed.

\substepseparator

\noindent\textbf{Identifying Important Inputs.} Before mutating inputs, we must first identify which of the inputs should be targeted. Real-world PLCs often process a large number of inputs, and attempting to include all of them may lead to challenges in scalability, especially when it comes to training: using a larger feature vector would require more data and could lead to a more complicated model.

To address this, we propose an \emph{information flow analysis} on the offline PLC code to determine which of the inputs are most \emph{important}, in the sense that they are most likely to have an effect on the output. In particular, we perform a dynamic dependency analysis that repeatedly mutates subsets of PLC inputs, executes the code, then increases an `importance score' for those inputs if the mutations changed the output. After several iterations, the most important inputs (i.e.~with the highest importance scores) can then be selected for constructing the feature vectors used in training.

\begin{algorithm}[!t]
\caption{Identifying the most important inputs}\label{alg:relevant_variables}
\KwIn{PLC code $C_P$; vector of $n$ inputs $I_{C_P}$; number of inputs to select $m$; set of labels $L$ encoding actuator commands}
\KwOut{Important inputs $I_{C_P}' \subseteq I_{C_P}$, where $|I_{C_P}'| = m$}
Initialise importance scores $S = [0, 0, \dots 0]$ where $|S| = n$;\\

\Repeat{timeout}
{
	Construct a vector $V = [v_0, v_1, \dots v_{n-1}]$ of values for $I_{C_P}$, extracted from a normal data trace;\\

	Let $l\in L$ denote the label encoding the actuator commands issued by $C_P$ on $V$;\\

	Construct $K$, a set of randomly selected indices $0 \leq k \leq n-1$;\\

	\For{$k \in K$}
	{
	Construct $V'$ from $V$ by randomly mutating the value of $v_k$;\\
	Let $l_k\in L$ denote the label encoding the actuator commands issued by $C_P$ on $V'$;\\

	\If{$l_k \neq l$}
	{
		$S[k] = S[k] + 1$;\\

	}
	}

}
Slice $I_{C_P}'$ from $I_{C_P}$ by selecting the $m$ inputs with highest importance scores in $S$;\\

\Return $I_{C_P}'$;
\end{algorithm}

Algorithm~\ref{alg:relevant_variables} summarises the steps of this process, which was inspired by the information flow analysis of Mathis et al.~\cite{Mathis-et_al17a}, but simplified to the setting of PLC programs. As the main body of PLC code is essentially a while-loop with a big switch-case structure, we do not instrument the code at intermediate points, but rather mutate different input values at random then observe and score the effects that these mutations have on the output. Again, this can be determined entirely by executing the offline PLC code, and does not require any process modelling.

Intuitively, Algorithm~\ref{alg:relevant_variables} works as follows. First, it instantiates the inputs with some values from a normal data trace (e.g.~extracted from a historian), and executes the offline code on them to determine the actuator commands that would be issued. Second, a subset of the values are mutated (independently and separately); if running the program again returns a different output, an importance score for the input is increased. Finally, after repeating the first two steps a number of times, the inputs with the highest importance score are returned. This simple and practical approximation can be used to select subsets of PLC inputs that are more likely to be useful for learning a model.

\emph{Implementation for SWaT.} The number of important inputs was manageable enough that all important inputs could be used in feature vectors (see Section~\ref{sec:evaluation}). That is, we used our analysis simply to identify the inputs that accumulated any importance score above zero. As a result, we included all sensor readings, actuator states, and variables concerning PLC state as input. We excluded variables that our analysis did not find to be important for our predictions, such as timers, and `healthy state' variables which only serve to raise alarms in the SCADA.

\substepseparator

\noindent\textbf{Generating Inputs.} Once the important input variables of a PLC have been identified, data collection can begin. We are aiming to train models on combinations of (important) inputs and the corresponding digital outputs of PLCs.  As the outputs are determined uniquely and deterministically by the PLC program, this is achieved simply by repeatedly generating multiple important input values at once, executing the offline PLC code on them, and logging the digital output commands issued. This is done without reference to the physical process, and thus can be repeated for any combination of valid inputs without the concerns of time (i.e.~due to slow processes), safety, or resource wastage. Furthermore, this approach is general and can be applied to a number of different CPSs, owing to the fact that process modelling is not required and that PLC programs are treated as black boxes.

In order to be able to train the model on sufficiently large data sets covering both normal and exception handling scenarios, and given that there is little cost to execute offline PLC code, we use a simple random mutation strategy to generate important input values. In general, there may be some heuristic strategies for generating important inputs, but they may lead to bias in the learnt models, as methods that influence the sampling probability lead to sampling bias. For example, we could use importance as the criterion of a heuristic strategy, but then the sampling probability of inputs with larger importance scores is higher, perhaps leading to a model that is less accurate on inputs with lower importance.

\emph{Implementation for SWaT.} For SWaT, we consider each offline PLC program in turn. The input values passed to each PLC consist of actuator states, the values of PLC variables, and discretised representations of sensor readings in the form of \emph{alarms}. For example, the continuous sensor reading of tank level sensor LIT101 can trigger four alarms---LIT101.H, LIT101.HH, LIT101.L, LIT101.LL---indicating four different height marks (High, HighHigh, Low, LowLow). We randomly generate several different combinations of inputs (alarms, actuators, variables), execute the offline PLC programs on them, and collect the digital actuator commands returned as output. We repeat this process until a certain amount of input/output data has been collected, as determined by experimentation (see Section~\ref{sec:evaluation}).

\subsection{Model Training and Validation}\label{sec:approach:step2}

In this next step, we use the data we collected to train models for predicting PLC behaviour, then validate that the predictions learnt from the offline code are also correct for the real PLCs.

\substepseparator

\noindent\textbf{Model Training.} With raw input/output data collected from our offline PLC programs, the next step is to organise that data into feature vectors that can be trained on by an appropriate learning algorithm (e.g.~neural network). In this work, we assume that the digital outputs of PLCs are actuator commands each indicating one of a finite number of discrete states that they should enter (e.g.~`actuator A123 ON'). With this assumption, we can learn multi-class single-label classifiers, in which the label represents a unique combination of discrete actuator states (i.e.~as commanded by the PLC outputs). Our feature vectors then consist of a fixed order of values for a number of important inputs with a single label representing the combination of actuator states that this would achieve. These are easily constructed from the raw input/output data collected in the previous step.

Once the feature vectors have been collected, we use them to train a supervised ML model. In particular, we explore neural networks. While certain other models (explored experimentally in Section~\ref{sec:evaluation}) are capable of learning the input/output relation, neural networks are an appropriate choice for our context as we can exploit their black box nature. First, they obfuscate the structure and workings of the PLC programs, ensuring that the attester maintains the `privacy' of that code (difficult to infer it from the neural network alone). Second, it is extremely difficult to attack both the PLC code and the neural network at the same time, in the sense of adding a wrong behaviour to the PLC and manipulating the neural network to consistently make that altered prediction. This gives an additional level of implicit trust that simpler models do not have.

\begin{figure}[!t]
	\centering
	\includegraphics[width=0.85\linewidth]{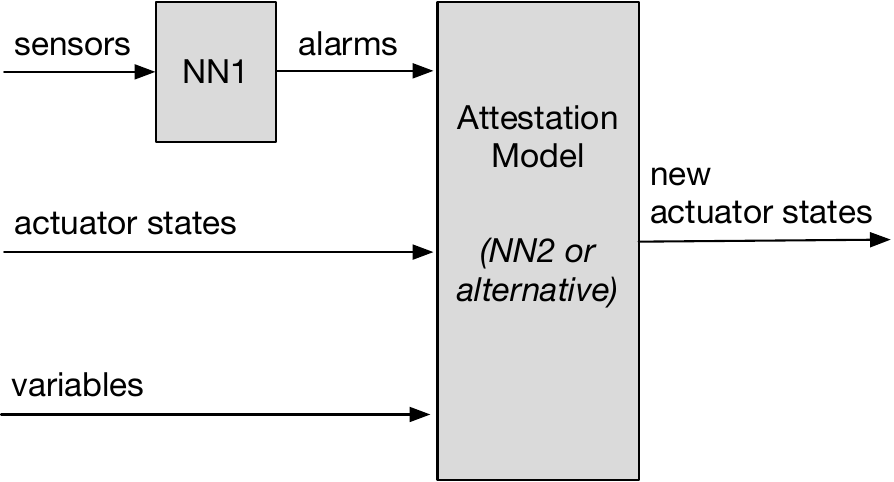}
	\caption{Behavioural attestation model for SWaT}
	\label{fig:attestation_model_swat}
\end{figure}

\emph{Implementation for SWaT.} The model required (Figure~\ref{fig:attestation_model_swat}) for SWaT is slightly more complicated as the PLC program expects not sensor readings, but rather alarm values (e.g.~LIT101.HH; Section~\ref{sec:approach:step1}). While it is possible to add some pre-processing code that maps sensor readings to alarms, given the presence of adversarial attackers, we instead chose to obfuscate this mapping by using a multi-label neural network (single layer; sigmoid activation for each node in the output layer). This neural network (identified as NN1 in Figure~\ref{fig:attestation_model_swat}) then feeds alarm values as inputs to the main (and more complicated) neural network we use for attestation, NN2. To train NN1, we use a simple program that randomly generates a large amount of sensor data and their associated alarms (the mappings of which are known in this case). We then train individual neural networks for single sensors, before combining them all into NN1, which is highly accurate (essentially $100\%$).

To train the main model, i.e.~the more complicated NN2, we organise the raw inputs into feature vectors and corresponding output labels:
\[ \langle [ NN1(s_0), NN1(s_1), \dots, a_0, a_1, \dots v_0, v_1, \dots ], l \rangle \]

\noindent where the vector contains a fixed order of alarm values $NN1(s_i)$, actuator states $a_i$, and variable values $v_i$. Note that for SWaT's PLCs, we include \emph{all} important sensors, actuators, and state variables (we exclude timers and healthy state variables, which are used to raise alarms to the SCADA system). Here, $l$ indicates a label that represents a unique combination of the actuator states relevant for the particular PLC. In PLC1, for example, there are three actuators: motorised valve MV101 and pumps P101/P102. We can uniquely represent the combination of their states using 4 bits: two bits for MV101 (which can be open, closed, or changing between the two), and one bit for each pump (which can be on or off). The label 14 (i.e.~1110), for example, represents the state in which pumps P101/P102 are on and valve MV101 is closed. These feature vectors are then used to train a neural network (NN2) with two hidden layers, containing 100 and 50 neurons respectively (ReLU as the activation function; Adam as the optimiser; learning rate of 0.001; cross-entropy loss function). In Section~\ref{sec:evaluation}, we also learn other kinds of models that could be used in conjunction to further build trust.

\substepseparator

\noindent\textbf{Model Validation.} Before utilising the learnt models as part of our remote attestation solution, it is important to validate that their predictions are also correct with respect to the \emph{real} system (a discrepancy could occur, for example, if there was a bug in the PLC code translation). To validate our models, we need to collect data by running the actual system under a range of normal conditions that span as many of the model inputs as possible. After this period of time, we extract the data logs from the system's historian, and perform a simple analysis: for every combination of inputs relevant to a PLC at time $t$, extract from the logs the states of relevant actuators at time $t+i$, then check that these are the same states as predicted by the models for those inputs. The value of $i$ is determined to reflect the maximum amount of time required for an actuator command to take effect in the system (as the state transition is typically not instant). Apart from checking the future states, we can compare the commands generated by the cyber part directly and immediately.

\emph{Implementation for SWaT.} We validated our models against data extracted from live runs of the real system, including sensor, actuator, and variable data arranged in a time series. We combined multiple runs of the system (approximately four hours in total) from different starting points, selected to maximise the dataset's coverage of normal states and thus ensure a larger number of combinations of PLC inputs. Our validation experiment (Section~\ref{sec:evaluation}) then proceeded as described in the general case, but with different values for $i$ to account for the different lengths of time it takes for a pump to change state (less than a second) and a motorised valve to fully open or close (approximately 8 seconds). Note that the actuator commands start taking effect \emph{immediately}, and these time delays are only to account for how long it takes for the actuators to complete their actions.

\subsection{Attesting the Behavioural Integrity of Code}

In this final step of our approach, we use our learnt models to build a remote attestation solution for the real PLCs. Intuitively, the idea is to constantly monitor the actual inputs/outputs of those PLCs, check them against the predictions of the models, and raise an alarm should the behaviour ever differ. Code attestation of this kind can be implemented in the workstation, or on independent devices (e.g.~Raspberry Pis) attached to the PLCs, that are able to read the inputs PLCs receive and the commands they send out. Our attestation scheme is simple yet practical for CPSs, as it can be implemented for any kind of PLC regardless of language or hardware. As we focus on modelling the control logic, we avoid the limitations of works (e.g.~\cite{Chen-Poskitt-Sun18a}) that assume the existence of accurate and comprehensive models of the physical processes.

While simple to implement, our approach maintains a high level of trust in the presence of powerful adversarial attackers with knowledge of the attesters and their underlying models. We exploit the black box nature of the underlying neural network to protect the privacy of the original PLC code, and rely on the fact that it is near-impossible to change the behaviour of the PLC code \emph{and} manipulate the neural network in such a way that it makes the same (wrong) prediction for that case. In addition, our attesting devices can run an ensemble of different predictive models simultaneously, such as multiple variants of neural networks, or a support vector machine~(SVM), in order to reduce the chances of adversarial attackers succeeding~\cite{Abbasi-Gagne17a,Tramer-et_al18a} and increase overall trust.

\emph{Implementation for SWaT.} Our behavioural attestation solution for SWaT follows exactly the aforementioned scheme; we expand upon details of the particular attestation models trained and attacks detected in our evaluation (Section~\ref{sec:evaluation}).

\section{Evaluation}\label{sec:evaluation}

We evaluate the effectiveness of our CPS attestation approach for detecting code modification attacks using its implementation for the SWaT testbed~(Section~\ref{sec:background}).

\subsection{Research Questions}

Our evaluation is designed around the following key research questions~(RQs):

\begin{description}
	\item[RQ1 (Training):] Which combination of data size and learning algorithm leads to the most accurate model?
	\item [RQ2 (Validation):] Is our attester able to effectively predict commands in a real system?
	\item [RQ3 (Attack Detection):] Is our attester able to detect code modification attacks?
	\item [RQ4 (Adversarial Attacks):] Is our attester robust against attackers with knowledge of the model?
\end{description}

RQ1 is concerned with the factors that lead to the most accurate model for predicting actuator commands from PLC inputs. These include the amount of data required and the variant of learning algorithm. RQ2 is concerned with validating that the learnt models of our attesters are capable of making accurate predictions for the real PLCs (and not just the offline translations of them). Finally, RQ3 and RQ4 are concerned with the effectiveness of our solution at fulfilling its primary purpose: detecting attacks. These include code modification attacks as well as adversarial attacks that exploit knowledge of the model's parameters in order to go undetected.

\subsection{Experiments and Discussion}

We present the design of our experiments for each RQ in turn, as well as the results and conclusions that we draw from them. The programs developed to implement these experiments on the SWaT testbed are all available online~\cite{Supplementary-Material}.

\substepseparator

\noindent\textbf{RQ1 (Training).} Our first RQ considers which combination of learning algorithm and training dataset size leads to the most accurate model for attestation.

To investigate this, based on the method described in Section~\ref{sec:approach}, we train models for all six of SWaT's PLCs using sets of data containing between 10,000 and 90,000 feature vectors at intervals of 10,000. The input values for each set is randomly generated afresh each time, i.e.~the 10,000 combinations of input values are not necessarily included in the 20,000 combinations of inputs values (etc.). In addition to neural networks, we also train SVM models and random forests~(RFs), to assess their potential use in an ensemble of attestation models. For each variant of learnt model, we calculate its accuracy using 5-fold cross validation, obtaining a ratio that indicates the number of correct predictions the classifier made for the feature vectors in the training set. A value of 0 indicates that all predictions were wrong, whereas 1 indicates that all predictions were correct.

\begin{table}[!t]
	\centering\footnotesize
	\caption{Accuracy \emph{(higher is better)} of the neural network attesters after training on different amounts of data}
	\label{tab:training_data_nn}
\begin{tabular}{c||c|c|c|c|c|c}
\#data & plc1 & plc2   & plc3   & plc4   & plc5   & plc6   \\ \hline\hline
10000  & 1    & 0.9905 & 0.9667 & 0.9660 & 0.9842 & 0.9888  \\ \hline
20000  & 1    & 0.9919 & 0.9937 & 0.9948 & 0.9988 & 0.9935  \\ \hline
30000  & 1    & 0.9938 & 0.9994 & 0.9984 & 0.9999 & 0.9972  \\ \hline
40000  & 1    & 0.9928 & 0.9999 & 0.9994 & 1      & 0.9982  \\ \hline
50000  & 1    & 0.9933 & 0.9999 & 0.9998 & 1      & 0.9990  \\ \hline
60000  & 1    & 0.9932 & 1      & 0.988  & 1      & 0.9998 \\ \hline
70000  & 1    & 0.9933 & 1      & 0.9999 & 1      & 0.9999   \\ \hline
80000  & 1    & 0.9927 & 1      & 1      & 1      & 0.9997\\ \hline
90000  & 1    & 0.9930 & 1      & 1      & 1      & 1

\end{tabular}
\end{table}

\begin{table}[!t]
	\centering\footnotesize
	\caption{Accuracy \emph{(higher is better)} of the SVM attesters after training on different amounts of data}
	\label{tab:training_data_svm}
\begin{tabular}{c||c|c|c|c|c|c}
\#data & plc1 & plc2   & plc3   & plc4   & plc5   & plc6   \\ \hline\hline
10000  & 0.9783 & 0.9865 & 0.9818 & 0.9267 & 0.9836 & 0.9873  \\ \hline
20000  & 0.9949 & 0.9885 & 0.9875 & 0.9528 & 0.9953 & 0.9970  \\ \hline
30000  & 0.9970 & 0.9910 & 0.9895 & 0.9609 & 0.9994 & 0.9998  \\ \hline
40000  & 0.9965 & 0.9900 & 0.9924 & 0.9694 & 0.9998 & 1    \\ \hline
50000  & 0.9974 & 0.9915 & 0.9942 & 0.9746 & 1      & 1  \\ \hline
60000  & 0.9976 & 0.9919 & 0.9965 & 0.9806 & 1      & 1 \\ \hline
70000  & 0.9981 & 0.9922 & 0.9975 & 0.9813 & 1      & 1   \\ \hline
80000  & 0.9981 & 0.9923 & 0.9989 & 0.9848 & 1      & 1\\ \hline
90000  & 0.9982 & 0.9917 & 0.9994 & 0.9868 & 1      & 1

\end{tabular}
\end{table}

\begin{table}[!t]
	\centering\footnotesize
	\caption{Accuracy \emph{(higher is better)} of the random forest attesters after training on different amounts of data}
	\label{tab:training_data_rf}
\begin{tabular}{c||c|c|c|c|c|c}
\#data & plc1 & plc2   & plc3   & plc4   & plc5   & plc6   \\ \hline\hline
10000  & 0.9724 & 0.9471 & 0.7147 & 0.8632 & 0.5272 & 0.9450  \\ \hline
20000  & 0.9843 & 0.9537 & 0.7336 & 0.8777 & 0.5355 & 0.9458  \\ \hline
30000  & 0.9843 & 0.9537 & 0.7336 & 0.8777 & 0.5355 & 0.9458  \\ \hline
40000  & 0.9885 & 0.9594 & 0.7294 & 0.8801 & 0.5103      & 0.9462  \\ \hline
50000  & 0.9881 & 0.9562 & 0.7221 & 0.8861 & 0.5378     & 0.9834  \\ \hline
60000  & 0.9873 & 0.9547 & 0.7281 & 0.8849 & 0.5441     & 0.9820 \\ \hline
70000  & 0.9842 & 0.9610 & 0.7337 & 0.8773 & 0.5447    & 0.9814   \\ \hline
80000  & 0.9840 & 0.9638 & 0.7310  & 0.8799 & 0.5490      & 0.9829\\ \hline
90000  & 0.9849 & 0.9612 & 0.7297 & 0.8798 & 0.5481     & 0.9841

\end{tabular}
\end{table}

\emph{Results.} Tables~\ref{tab:training_data_nn}, \ref{tab:training_data_svm}, and \ref{tab:training_data_rf} present the results of this experiment. Ultimately, accuracy is very high across the board for our neural network and SVM models, indicating that the models are expressive enough to accurately predict the outputs of PLCs for every process of SWaT. The RF models, however, suffer from lower accuracies for PLCs 2--5, which we suspect is due to their sensitivity to large numbers of features, as well as the larger number of PLC state variables; in particular, 20 such variables in PLC5. While our NN and SVM models are already very accurate for training sets of size 10,000, we choose to fix 90,000 as our amount of training data for the remaining experiments to maximise the accuracy. Given that the data is collected entirely from the offline PLC code (i.e.~no process simulation involved), it can be collected very quickly.

\begin{center}
\noindent\fbox{%
    \parbox{0.75\linewidth}{%
        \small\emph{Our neural network attestation model achieves near-$100\%$ accuracy across all data set sizes.}
    }
}
\end{center}

\substepseparator

\noindent\textbf{RQ2 (Validation).} Although our first RQ established that our neural network models are highly accurate, this is based on training data obtained from the offline PLC programs. In this RQ, we aim to consider whether the accuracy of the models also translates to the \emph{real} PLCs of the CPS.

To investigate this, we performed multiple runs of the SWaT testbed and collected a comprehensive dataset consisting of sensor, actuator, and variable values. Rather than run the system continuously, we initialised multiple runs from different starting configurations, in order to maximise a simple coverage criterion: to cover \emph{all normal states} of SWaT. We collected 4 hours of data from this process, covering all normal states and thus maximising the number of different PLC inputs to validate our models against. For each row of data, we extracted the inputs and fed them to our learnt model, then compared the actuator states predicted against the \emph{actual} actuator states logged after a number of seconds. In particular, we checked after 1s for pumps, whereas for valves, which change state more slowly, we checked for predicted changes between 7-10s after the command is issued. For each model, we calculated the \emph{false alarm rate}, i.e.~a ratio expressing the number of wrongly predicted actuator states (as SWaT was not under attack, all such wrong predictions would be considered false attestation alarms). A value of 0 indicates that there were no false alarms; a value of 1 indicates that every prediction was considered a false alarm.

\emph{Results.} Table~\ref{tab:false_alarms} presents our results for the PLC attesters across all variants. Our neural network and SVM models perform very well, with (near-)zero false predictions for PLCs 1, 3, 5, and 6. The neural network outperformed the SVM model on PLC4, seeing false alarm rates of 0.0001 and 0.1219 respectively. The RF model, interestingly, had some success for PLC4 (0.0001), but in most cases performed much more poorly than the other models. Using the RF would be ineffective in practice, due to the large number of false alarms raised.

\begin{table}[!t]
	\centering\footnotesize
	\caption{False alarm rates \emph{(lower is better)} for different attestation models (before retraining)}
	\label{tab:false_alarms}
\begin{tabular}{c||c|c|c|c|c|c}
Model      & plc1    & plc2   & plc3   & plc4   & plc5   & plc6   \\ \hline\hline
NN         & 0.0009	 & 0.239  &	0.004  & 0.0001	&0	&0  \\ \hline
SVM        & 0.0009	&0.2395	&0.004	&0.1219	&0	&0  \\ \hline
RF         & 0.1108	&0.2385	&0.323	&0.0001	&0.5388	&0.033  
\end{tabular}
\end{table}

Despite the near-zero false alarm rates for PLCs 1 and 3--6, the neural network (as well as the SVM) model performed surprisingly poorly for PLC2 (0.239). Upon investigation, we found that the model was predicting just one combination of inputs incorrectly, and that this particular combination appeared several times in the dataset (i.e.~bias). We added 100 copies of the correctly labelled feature vectors to the training set, retrained the neural network, and were able to achieve a new false alarm rate for PLC2 of 0.0001 (0.0002 for SVM; 0.0001 for RF).

\begin{center}
\noindent\fbox{%
    \parbox{0.75\linewidth}{%
        \small\emph{Following its re-training, our neural network attestation model can predict future actuator states with practically no false positives.}
    }
}
\end{center}

\substepseparator

\noindent\textbf{RQ3 (Attack Detection).} Our third RQ is concerned with the utility of our attestation solution: can it actually detect attacks in practice? In particular, can it detect \emph{code modification attacks} on the PLCs? Given that we are attesting the behavioural integrity of PLC code, we refine this question further to whether attacks that change the input/output behaviour of the PLCs---in any way---can be detected?

To investigate this question, we designed a similar experiment to that of our previous work~\cite{Chen-Poskitt-Sun18a}, and randomly generated mutants for every PLC program using their faithful offline translations. We randomly generated 20 mutants for each PLC, using the same mutation operators as~\cite{Chen-Poskitt-Sun18a} to generate one mutation per copy. Furthermore, we also had to ensure that our mutants were \emph{effective}. In our previous work, this was done by running the mutated code on a simulated physical process. In our new approach, we only need the (fully accurate) PLC code to establish effectiveness. In particular, given a mutant program and the original, we run both with 5000 randomly generated inputs until we uncover different actuator outputs, establishing the effectiveness of the given mutant (if no differences in output are found, the mutant is discarded and replaced). For each set of 20 effective mutants, we then randomly generate inputs until we have 1000 that produce different outputs between the mutant and the original. Finally, for each of these inputs, we use our attestation model to predict what the (normal) output should be for the given inputs, and deem the attack \emph{detected} if the prediction is different from the outputs caused by the mutant.

\emph{Results.} We used our neural network (and SVM) attestation models to make predictions for all 120,000 combinations of inputs (i.e.~1000 effective input combinations for each of the 120 mutants). We found that both the models could \emph{detect all mutants} across all the input combinations (i.e.~the predictions differed from the abnormal outputs of the mutants), suggesting their efficacy for detecting code modification attacks. As our RF model had high false alarm rates (RQ2), we could not meaningfully evaluate it as a detector.

\begin{center}
\noindent\fbox{%
    \parbox{0.75\linewidth}{%
        \small\emph{Our PLC code attesters were able to detect all 120 code modification attacks with $100\%$ success.}
    }
}
\end{center}

\substepseparator

\noindent\textbf{RQ4 (Adversarial Attacks).} Our fourth and final RQ assesses the robustness of our models against adversarial attackers, i.e.~attackers that have knowledge of the model's parameters, and can craft noise designed to cause it to make an incorrect prediction (e.g.~to mask a code modification in the actual PLC). This experiment is important as recent work (e.g.~\cite{Papernot-et_al16a,Wang-et_al19a,Ghamizi-et_al20a,Jia-et_al21a}) suggests that neural networks are brittle against adversarial samples. We investigate the RQ in two parts: (1)~is the adversarial attacker able to achieve \emph{any} change of behaviour within various bounds of noise; and (2)~is the adversarial attacker able to achieve a \emph{specific} change of behaviour within those same bounds of noise?

To investigate these questions, first, the attacker must tackle NN1 (Figure~\ref{fig:attestation_model_swat}), the neural network that maps sensor readings to alarms. Essentially, the attacker applies small amounts of noise (within bounds of $1\%$, $5\%$, and $10\%$) to 20,000 combinations of inputs extracted from the normal dataset~\cite{Goh-et_al16a}, in order to compute which alarms it is capable of changing. Note that any amount of noise larger than this is very likely to be detected by other defence mechanisms in the system. Then, for part (1), the attacker searches for a combination of these alarms that causes one of the actuator commands in NN2's output to change, marking a success if such a combination can be found. For part (2), we randomly sample 20,000 input values from the normal dataset, and compute the outputs that NN2 will give for each of them. Next, we `flip' one of the actuator commands in these computed outputs, and then have the adversarial attacker search for alarms that it can modify (within the noise bound) that bring about that same actuator output. This essentially represents a scenario where an attacker is attempting to cause a specific (mis-)prediction so as to mask a separate PLC code modification attack.


\emph{Results.} Table~\ref{tab:adversarial_attacks} presents the \emph{success rates} for an adversarial attacker engaging in the first scenario described earlier. Again, given a set of inputs, success is declared if the attacker is able to determine the alarms it can manipulate by applying $n\%$ of noise to sensor inputs and use those alarms to change \emph{any} of the outputs of NN2. As can be seen, for both our neural network (and SVM) models, the attacker only becomes successful at effecting change for the higher bounds of noise. These are large levels of noise which are likely to be detected by other standard defence mechanisms in the system, e.g.~anomaly detectors or invariant checkers.

\begin{table}[!t]
	\centering
	\caption{Success rates: adding noise to achieve \emph{any} change of output for NN2}
	\label{tab:adversarial_attacks}
\begin{tabular}{c||c|c|c}
Noise     & 1\%  & 5\%   & 10\% \\ \hline\hline
NN         & 0.094 &	0.5354 &	0.61 \\ \hline

SVM        & 0.1061 &	0.6817 &	0.7651  \\ 
\end{tabular}
\end{table}

Table~\ref{tab:adversarial_attacks_particular} presents the success rates for an adversarial attacker engaging in the second scenario. This time, success is declared if the attacker is able to determine the alarms it can manipulate by applying $n\%$ of noise to sensor inputs and use those alarms to change the outputs of NN2 to some \emph{specific} output. In comparison to the first scenario, the success rates of the attackers plunge (e.g.~$0.0018$ for neural networks using $1\%$ noise), and the numbers remain very low even for higher levels of noise (less than $0.02$ for neural networks). These numbers illustrate the challenge an attacker would face in their goal to change the behaviour of a PLC and then also manipulate the attestation model to make a consistent (mis-)prediction. Even if such a powerful attacker existed, achieving a desired and consistent manipulation within a reasonable amount of time is unlikely to be practical.

\begin{table}[!t]
	\centering
	\caption{Success rates: adding noise to achieve a \emph{particular} change of output for NN2}
	\label{tab:adversarial_attacks_particular}
\begin{tabular}{c||c|c|c}
Noise     & 1\%  & 5\%   & 10\% \\ \hline\hline
NN         & 0.0018	&0.0143	&0.0159 \\ \hline
SVM        & 0.0075 &	0.0190 &	0.0255 \\ 
\end{tabular}
\end{table}
\begin{center}
\noindent\fbox{%
    \parbox{0.75\linewidth}{%
        \small\emph{It is not practically possible for an adversarial attacker to consistently manipulate the PLC code and the model using discreet levels of noise.}
    }
}
\end{center}

We remark that according to our threat model, the attacker could also potentially change one of the actuator values fed as input to NN2. However, this would be caught immediately as actuator values are discrete (as opposed to continuous sensor values, in which noise can affect the neural network but not the PLC).

\subsection{Threats to Validity and Limitations}

There are some threats to the validity of our results. First, as our experiments were applied to a single CPS, it is possible that the results do not generalise to other systems. However, we highlight that our approach was designed with generality as a principal goal: unlike previous attestation work (e.g.~\cite{Chen-Poskitt-Sun18a}), we do not require any process modelling to train our attestation models: we only need faithful offline copies of the PLC code. For SWaT, we translated and learnt from programs written in an industry-standard language---structured text---suggesting that the PLCs of other systems should be possible to translate and train on too.

Second, owing to the lack of benchmarks for code modification attacks on CPSs, and the lack of documented code modification attacks for SWaT specifically, we followed our previous work~\cite{Chen-Poskitt-Sun18a} in evaluating the attesters on randomly generated code modification attacks (i.e.~mutants). It is possible that these mutants are not representative of the modifications that an intelligent attacker might make, and that our results might not apply to certain zero-day attacks. However, the high levels of accuracy, $100\%$ success rates for random (effective) code modifications, and the solution's robustness against consistent adversarial attacks increase our confidence.

The main limitation of our approach is that code modifications are only detected at the moment they impact the black box input/output behaviour of a PLC. Unlike traditional hardware and software attestation approaches (Section~\ref{sec:background}), we would not be able to to detect a modification that has no impact on the PLC's behaviour (e.g.~the equivalent of `skip'). In practice, however, PLC attackers must change the input/output behaviour to eventually be able to cause physical damage elsewhere in the system. Our approach can detect an incorrect command the moment it occurs and immediately issue a warning.

\section{Related Work}\label{sec:related_work}

In this section, we highlight some related work addressing the broader themes of this paper: ensuring the integrity of CPSs and critical infrastructure in particular. Note that key related work specifically addressing the remote attestation of embedded devices was reviewed earlier, in Section~\ref{sec:background}.

Several different approaches have emerged over the last few years for detecting and preventing CPS attacks. These include techniques based on \emph{anomaly detection}, where the logs of the physical data are analysed to identify suspicious events and anomalous behaviours~\cite{Cheng-Tian-Yao17a,Harada-et_al17a,Inoue-et_al17a,Pasqualetti-Dorfler-Bullo11a,Aggarwal-et_al18a,Aoudi-et_al18a,He-et_al19a,Kravchik-Shabtai18a,Lin-et_al18a,Narayanan-Bobba18a,Schneider-Boettinger18a,Carrasco-Wu19a,Kim-Yun-Kim19a,Adepu-et_al20a,Das-Adepu-Zhou20a,Giraldo-et_al20a,Schmidt-Hauer-Pretschner20a}; digital fingerprinting, where sensors are checked for spoofing by monitoring time and frequency domain features from sensor and process noise~\cite{Formby-et_al16a,Gu-et_al18a,Kneib-Huth18a,Ahmed-et_al20a,Yang-et_al20a}; and invariant-based checkers, which monitor for violations of invariant properties over sensor and actuator states~\cite{Cardenas-et_al11a,Adepu-Mathur16a,Adepu-Mathur16b,Chen-Poskitt-Sun16a,Adepu-Mathur18b,Chen-Poskitt-Sun18a,Choi-et_al18a,Giraldo-et_al18a,Umer-et_al20a,Yoong-et_al21a}. A key difference between many of these works and ours, aside from the specific target of code modification attacks, is the need for either significant amounts of data from the real system (for training), or knowledge of the underlying processes and control operations (e.g.~for designing physical invariants). Ultimately, we see these approaches as complimentary to our code integrity checking approach: fingerprinting approaches, for example, might complement our behavioural attesters by detecting adversarial noise in the network (Section~\ref{sec:evaluation}).

The strengths and weaknesses of different countermeasures has been the focus of various studies. Erba and Tippenhauer~\cite{Erba-Tippenhauer20a} spoof sensor values (e.g.~using precomputed patterns) and are able to evade three black box anomaly detectors published at top security conferences, highlighting the need for complementary approaches such as ours. Urbina et al.~\cite{Urbina-et_al16a} evaluated several attack detection mechanisms in a comprehensive review, concluding that many of them are not limiting the impact of stealthy attacks. Our code integrity checking solution addresses this point in the sense that we assume attackers may have knowledge of the attestation model, then exploit the black box nature of the neural network in order to maintain trust. C\'{a}rdenas et al.~\cite{Cardenas-et_al14a} propose a general framework for assessing attack detection mechanisms, but in contrast to the previous works, focus on the business cases between different solutions. For example, they consider the cost-benefit trade-offs and attack threats associated with different methods, e.g.~centralised vs.~distributed. Sun et al.~\cite{Sun-et_al20a} systemise existing knowledge on PLC code modification attacks specifically, concluding that PLCs need a full chain of protection where solutions such as ours are used in tandem with complementary approaches, e.g.~based on formal verification (e.g.~\cite{Wang-et_al21a}) or stealthy attack detection.

The need to develop a chain of protection for PLCs is further motivated by the emergence of advanced attacks that target these devices specifically~\cite{Basnight-et_al13a}. Garcia et al.~\cite{Garcia-et_al17a}, for example, demonstrate a feasible attack within the firmware of Allen Bradley PLCs, using a physics-aware rootkit that can modify control commands before they are sent out to actuators. PLCs can also fall prey to worms, such as PLC-Blaster~\cite{Spenneberg-et_al16a}, which injects malicious control logic into vulnerable PLCs identified after a network scan. Our PLC code attestation solution is specifically designed to detect such changes in behaviour, but will only do so once the injected code is executed and the actuator commands issued are changed. This defence strategy could be complemented by a number of other PLC-compatible approaches, e.g.~dynamically generating/monitoring control invariants from SCADA logs~\cite{Yang-et_al20b}, fingerprinting~\cite{Ahmed-et_al21a}, mining temporal invariants/dependencies for safety vetting~\cite{Zhang-et_al19a}, or control flow integrity monitoring~\cite{Abbasi-et_al17a}.

The idea of using code mutations as a way to obtain and learn from abnormal behaviour of CPSs was first explored in~\cite{Chen-Poskitt-Sun16a,Chen-Poskitt-Sun18a}. Instead of mutating the PLC code, our work mutates the PLC inputs, leading to a more general and practical attestation solution. Mutations have been explored in a number of other CPS contexts too. For example, Brandl et al.~\cite{Brandl-Weiglhofer-Aichernig10a} apply mutations to specifications of hybrid systems (rather than to the PLC programs themselves) in order to derive distinguishing model-based test cases that can be seen as classifiers. A discrete view of the system is used for generating test cases, with qualitative reasoning applied to represent the continuous part. Chowdhury et al.~\cite{Chowdhury-et_al20a} also use mutations, but to find bugs in model-based design tools for CPSs, such as Simulink.

As a testbed dedicated for cyber-security research, many different countermeasures have been developed for SWaT itself. These include anomaly detectors, typically trained on the publicly released dataset~\cite{Goh-et_al16a,CPS-Datasets} using unsupervised learning techniques, e.g.~\cite{Goh_et-al17a,Inoue-et_al17a,Kravchik-Shabtai18a}. Ahmed et al.~\cite{Ahmed-et_al20a} implemented fingerprinting systems based on sensor and process noise for detecting spoofing. Adepu and Mathur~\cite{Adepu-Mathur16a,Adepu-Mathur16b,Adepu-Mathur18b} systematically and manually derived physics-based invariants and other conditions to be monitored during the operation of SWaT. Feng et al.~\cite{Feng-et_al19a} also generate invariants, but use an approach based on learning and data mining that can capture noise in sensor measurements more easily than manual approaches.

\section{Conclusion}\label{sec:conclusion}

We proposed a \emph{practical code integrity checking solution} based on privacy-preserving black box models that attest the behaviour of PLC programs. Our approach advances the state-of-the-art through its generality and practicality: by focusing on translations of the PLC code instead of process models as in previous work~\cite{Chen-Poskitt-Sun18a}, our models can be trained for real critical infrastructure---not just process simulators---entirely offline, thus avoiding the problems of resource wastage and safety concerns. Furthermore, our code integrity checking scheme can be implemented without requiring any firmware modification or trusted hardware components (problematic for proprietary and legacy PLCs), and by exploiting the black box nature of the underlying neural network, does not assume that attackers are unaware of its presence or model parameters. We implemented our technique for SWaT, a six-stage water treatment testbed, learning attestation models that could achieve near-$100\%$ accuracy and detect all 120 of a series of PLC code modification attacks. Finally, we subjected our attesters to the manipulations of an adversarial attacker, finding that it was not practically possible to change the behaviour of the PLCs while simultaneously causing the attesters to make consistent (mis-)predictions using only discreet levels of adversarial noise.

\begin{acks}
	We are grateful to the anonymous referees for their insightful reviews and suggestions, which have helped to improve the quality of this paper. This research / project is supported by the National Research Foundation, Singapore, under its National Satellite of Excellence Programme ``Design Science and Technology for Secure Critical Infrastructure'' (Award Number:~NSoE\_DeST-SCI2019-0008). Any opinions, findings and conclusions or recommendations expressed in this material are those of the author(s) and do not reflect the views of National Research Foundation, Singapore.
\end{acks}

\bibliographystyle{ACM-Reference-Format}
\balance
\bibliography{references}

\end{document}